\documentclass[aps,twocolumn,nofootinbib,showpacs]{revtex4}
\usepackage{natbib}
\usepackage{amssymb,amsbsy,amsmath,amsfonts}
\usepackage{graphicx}
\usepackage{float}
\usepackage{rotating}

\begin{document}
\title {Role of 
the $N^*$(1535) in the $J/\psi\rightarrow \bar{p}\eta p$ and 
$J/\psi\rightarrow \bar{p}K^+\Lambda$ reactions}

\author{L. S. Geng,$^1$ E. Oset,$^1$ B. S. Zou,$^2$ and M. D\"oring$^3$}
 \affiliation{ 
 $^1$Departamento de F\'{\i}sica Te\'orica and IFIC, Universidad de Valencia-CSIC,
E-46071 Valencia, Spain\\
 $^2$ Institute of High Energy Physics, Chinese Academy of Sciences, \\
 P.O.Box 918(4), Beijing 100049, China\\
 $^3$Institut f\"ur Kernphysik, Forschungszentrum J\"ulich GmbH, 52425 J\"ulich, Germany}

\begin{abstract}
We study the $J/\psi\rightarrow \bar{p}\eta p$ and $J/\psi\rightarrow \bar{p}K^+\Lambda$
reactions with a unitary chiral approach. We find that the unitary chiral approach, which 
generates the $N^*(1535)$ dynamically, can describe the data reasonably well, particularly the ratio of
the integrated cross sections. This study  provides further support
for the unitary chiral  description of the $N^*(1535)$. 
We also discuss some subtle differences between the coupling constants determined from the unitary chiral approach and those determined from phenomenological studies. 

\end{abstract}
\date{\today}
 \maketitle

\section{Introduction}
Understanding the nature of various hadrons has always been a main 
goal pursued in studies of strong interaction phenomena. With the advent of
quantum chromodynamics (QCD) the hope was raised that one could understand
the various hadrons observed in nature as
quarks and gluons bound together by
the strong interaction. For instance, in
terms of these degrees of freedom,
baryons and mesons are often seen as $qqq$ or $q\bar{q}$ composites, respectively. There are, however, certain resonances that cannot easily fit into
this picture, for instance, the $\Lambda(1405)$ and the $N^*(1535)$.

The $N^*(1535)$ with a mass higher than that of the lowest $J^P=1/2^+$ radial excitation state 
$N^*(1440)$ has long been a problem in conventional quark models~\cite{Capstick:2000qj}.
In recent years, a new interpretation has been proposed based on studies performed within unitary chiral theories (U$\chi$PT); i.e., 
it is dynamically generated from the interaction of the octet of the pseudoscalar mesons and the octet 
of the proton~\cite{Kaiser:1995cy,Kaiser:1996js,Nieves:2001wt,Inoue:2001ip}. In these studies,
its extremely strong coupling to the $\eta N$ channel~\cite{Yao:2006px} comes out naturally. In addition,
a strong coupling of the $N^*(1535)$ to
the $K\Sigma$ and $K\Lambda$ channels is predicted, the latter
seems to be consistent with recent analyses of the $J/\psi\rightarrow \bar{p}K^+\Lambda ~$\cite{Liu:2005pm,Liu:2006ym},
$pp\rightarrow p K^+\Lambda$~\cite{Liu:2006tf}, and $\gamma p\rightarrow K^+\Lambda$ reactions~\cite{Shklyar:2005xg,JuliaDiaz:2006is}. Several further studies
utilizing the U$\chi$PT amplitudes have also been performed recently~
\cite{Doring:2005bx,Jido:2007sm,Doring:2008sv}, which all support the
U$\chi$PT description of the $N^*(1535)$.

The $J/\psi$ and $\psi'$ experiments at the Beijing Electron-Positron Collider (BEPC)
provide an excellent place for studying excited nucleons and hyperons~\cite{Zou:2000wg}.
In Ref.~\cite{Liu:2005pm}, based on the BES results on $J/\psi\rightarrow\bar{p}\eta p$~\cite{Bai:2001ua} and
$J/\psi\rightarrow 
\bar{p}K^+\Lambda$~\cite{Yang:2005ej}, the ratio between the 
effective coupling constants of the $N^*(1535)$
to $K\Lambda$ and $p\eta$ is determined to be $R=g_{N^*(1535)K\Lambda}/g_{N^*(1535) \eta p}=1.3\pm0.3$. Together
with the previously fixed $g_{N^*(1535) \eta p}$, they were able to reproduce recent 
$pp\rightarrow p K^+\Lambda$ near-threshold cross-section data~\cite{Balewski:1998pd,Bilger:1998jf,Sewerin:1998ky,Kowina:2004kr} very well.

In Ref.~\cite{Liu:2005pm}, it was noted that the $g_{N^*(1535)K\Lambda}/g_{N^*(1535) \eta p}$
ratio obtained there by fitting the BES data is larger by a factor of two than the corresponding
U$\chi$PT one~\cite{Inoue:2001ip}. This raises naturally the question 
whether the U$\chi$PT picture of the $N^*(1535)$ is consistent with the BES data and how to
understand the difference in the values of the coupling constants.
In the present work, we aim to answer these questions by studying 
the reactions $J/\psi\rightarrow \bar{p}\eta p$
and $J/\psi\rightarrow \bar{p}K^+\Lambda$ within the unitary chiral approach.

This article is organized as follows. In Sec. II, we briefly outline
the unitary chiral theory and the dynamical generation of the $N^*(1535)$.
In Sec. III, we lay down the formalisms to study the reactions $J/\psi\rightarrow
\bar{p}\eta p$ and $J/\psi\rightarrow \bar{p}K^+\Lambda$.  Results and discussions are
given in Sec. IV, followed by a brief summary in Sec. V.

\section{Unitary chiral theory and the dynamical generation of the $N^*(1535)$}
Unitary chiral theories start with an interaction kernel, $V$, provided by the corresponding
chiral Lagrangians, either lowest order or higher order. 
In Ref.~\cite{Kaiser:1995cy} the Lippmann-Schwinger equation in coupled channels was used to
provide a unitary amplitude in the study of meson-baryon interaction. In Ref.~\cite{Oller:1997ti}
also the Lippmann-Schwinger equation in coupled channels was used in the case of the meson-meson
interaction. Yet, as noted in Ref.~\cite{Lee:1998gt}, the method of Ref.~\cite{Oller:1997ti}, integrating
explicitly the $q^0$ variable in the loops and using relativistic propagators, corresponds to
a coupled channel Bethe Salpeter equation, and most of the recent works on the topic
\cite{Nieves:2001wt,Inoue:2001ip,Hyodo:2002pk,Oset:2001cn,Borasoy:2005ie,Oller:2006yh,Borasoy:2005du,Borasoy:2006sr} adhere to this method and this nomenclature.

Other unitarization procedures are obtained using the Inverse Amplitude Method (IAM)~\cite{Truong:1988zp,Guerrero:1998ei,
Pelaez:2004xp}
and  the $N/D$ method~\cite{Oller:1998zr,Oller:2000fj}.

In the Bethe-Salpeter equation method, which we employ in the present work,
one has in matrix form
\begin{equation}\label{eq:bethe}
T=(1-VG)^{-1}V,
\end{equation}
where $T$, $V$ are complex matrices in coupled channels and $G$ is a diagonal matrix
with its element the two-body loop function. 
In ``full form,'' the Bethe-Salpeter equation is an integral equation where the kernel
$V$ has the full spin and angular momentum dependence and the propagators appear in their full
covariant form (see, e.g., Refs.~\cite{Salpeter:1951sz,Lahiff:1999ur}).
In the present case, one studies only the $s$-wave scattering amplitude and
$V$ is already projected in $s$ wave. In addition, in the
case of meson-baryon interaction, only the positive energy part of the baryon
propagator  (with relativistic energies) is kept, while the relativistic propagator of the
mesons is taken. As it has been demonstrated with numerous examples,
one can render the complex integral equations  into 
algebraic ones by using the on-shell approach with the argument that the off-shell components can be absorbed by redefining
the corresponding coupling constants~\cite{Oller:1997ti}. It also finds an equivalent interpretation in the 
$N/D$ method that relies upon a dispersion relation for $T^{-1}$~\cite{Oller:1998zr,Oller:2000fj}.

The equivalence of the $N/D$ method and the on-shell factorized Bethe-Salpeter equation, 
Eq.~(\ref{eq:bethe}), follows when using the $N/D$ method, neglecting the left-hand cut as a source of
the imaginary part in the dispersion relation (see Ref.~\cite{Oset:2005ag} for a precise and pedagogical exposition). As described in Refs.~\cite{Oller:1998zr} and \cite{Oller:2000fj}, the contribution of the left-hand cut in the physical
region is either very small or, in any case, very weakly energy dependent, such that its
effects are easily incorporated by means of the subtraction constants of the dispersion integral.
A more detailed explanation of these facts can be found in Sec. II of Ref.~\cite{Doring:2006ub}.

To study the $J/\psi\rightarrow \bar{p}\eta p$ and $J/\psi\rightarrow
\bar{p}K^+\Lambda$ reactions through intermediate $N^*(1535)$ [$\bar{N}^*(1535)$], we are interested in the $S=0$ and $Q=+1$ sector with
the following six coupled channels:
\begin{equation}
 \pi^0p,\;\pi^+n,\;\eta p,\;K^+\Sigma^0,\; K^+\Lambda,\;K^0\Sigma^+.
\end{equation}

The lowest order chiral Lagrangian responsible for the meson-baryon interaction is~\cite{Bernard:1995dp}
\begin{eqnarray}\label{eq:lag}
 \mathcal{L}
 &=&\langle\bar{B}(i\gamma^\mu D_\mu -M_B)B\rangle\nonumber\\
&&+\frac{D}{2}\langle\bar{B}\gamma^\mu\gamma_5[u_\mu,B]\rangle
+\frac{F}{2}\langle\bar{B}\gamma^\mu\gamma_5\{u_\mu,B\}\rangle.
\end{eqnarray}
The term with the covariant derivative, $D_\mu$, in this Lagrangian provides the $MMBB$ transition amplitude, i.e., the
Weinberg-Tomozawa interaction,
\begin{equation}\label{vij}
 V_{ij}=-C_{ij}\frac{1}{4f_i f_j}\bar{u}(p')\gamma^\mu u(p)(k_\mu+k'_\mu)
\end{equation}
with $p$ ($k$), $p'$ ($k'$) being the initial and final momenta of the baryons (mesons).
The coefficients $C_{ij}$ can be found in Ref.~\cite{Doring:2005bx}. 
The terms with the $D$ and $F$ couplings account for the Yukawa coupling of a single meson to
baryons and will play a role, by analogy, in the posterior discussions.
At low energies, the amplitudes $V_{ij}$ can be simplified by retaining the largely dominant $\gamma^0$ component and one finds
an easy analytical expression for $V_{ij}$~\cite{Inoue:2001ip,Oset:2001cn}.
In Eq.(\ref{vij}) $f_i$ ($f_j$) is
the meson decay constant with $f_\pi=93$ MeV, $f_K=1.22 f_\pi$, and $f_\eta=1.3 f_\pi$~\cite{Inoue:2001ip}.

The use of different meson decay constants, as well as other details of the calculation
in Ref.~\cite{Inoue:2001ip} require some explanations. In Ref.~\cite{Inoue:2001ip} only the lowest order meson-baryon
chiral Lagrangian of Eq.(\ref{eq:lag}) is used. The subtraction constants in the
dispersion integral or loop function, $G$ of Eq.(\ref{eq:bethe}), are assumed to account for effects
of higher order Lagrangians. There is, however, a caveat in this assumption because in
chiral perturbation theory the loop terms contribute to order $Q^3$ for the case of meson-baryon interaction 
(the counting is different in the meson-meson interaction), while there are chiral Lagrangians of
order $Q^2$ that would not be accounted for by means of the subtraction constants~\cite{Kaiser:1996js,Borasoy:2005ie,Oller:2006yh,Borasoy:2006sr}. The effects of using different
$f_i$ are also technically of order $Q^3$. Although the unitary resummation will mix 
different powers of $Q$, the aim of the chiral unitary approach is to provide a unitary framework
at higher energies that matches exactly the chiral perturbation theory amplitude
at low energies~\cite{Oller:2000fj}. For the meson-baryon interaction the matching should be done at order $Q^3$.
The ability of the method used in Ref.~\cite{Inoue:2001ip} to provide realistic amplitudes depends 
upon the $Q^2$ terms being small. This is, of course,  a matter of principle. In practice,
and as one is usually concerned about a relatively narrow band of energies, the subtraction
constants can approximately account for these $Q^2$ terms. Ultimately, it is the
comparison of theoretical calculations done with the lowest order Lagrangian with those including higher order
terms that
must tell us how accurate the lowest order can be. Such a comparison is possible now.
Indeed, in Ref.~\cite{Borasoy:2006sr}, where higher order Lagrangians are used, an estimation of theoretical errors
is done. This is very useful and, comparing the results obtained there with those of Ref.~\cite{Oset:1997it} using
only the lowest order Lagrangian, one can see that the results with the lowest
order  fall well within the theoretical uncertainties of the higher order calculations.

Two modifications to the above transition amplitudes of Eq.~(\ref{vij}) must be introduced to
better describe the phase shifts and inelasticities of $S_{11}$ and $S_{31}$ $\pi N$ scattering.
The first modification is due to the realization that the lowest order chiral Lagrangian may be
viewed as an effective manifestation of the vector-meson exchange between the mesons and the
baryons in an alternative picture, the hidden gauge formalism~\cite{Bando:1984ej,Bando:1987br}, which is shown to
be equivalent to the use of chiral Lagrangians~\cite{Ecker:1989yg}. Therefore,
to account for the dependence on the momentum transfer of the vector-meson propagator, one
replaces $C_{ij}$ with
\begin{equation}
 C_{ij}\int\frac{d \hat{k}'}{4\pi}\frac{-m^2_v}{(k'-k)^2-m_v^2}
\end{equation}
at 
\begin{equation}
 \sqrt{s}>\sqrt{s^0_{ij}},
\end{equation}
where $\sqrt{s^0_{ij}}$ is the energy where the above integral is unity,
and which appears between the thresholds of the two $i$, $j$ channels.

The second modification is the effective inclusion of the $\pi N\rightarrow \pi\pi N$ channel.
This channel was very important to obtain a good description of the $I=3/2$ amplitudes
but it has only a small influence in the $I=1/2$ channel~\cite{Inoue:2001ip}.
Following Refs.~\cite{Doring:2005bx} and \cite{Doring:2004kt}, in the $Q=+1$ sector,
this can be achieved by a modification of the potential, i.e.,
$V_{\pi N\rightarrow\pi N}\rightarrow V_{\pi N\rightarrow\pi N}+\delta V\times G_{\pi\pi N}$, with
$\delta V$ given by
\begin{eqnarray}
 \delta V_{\pi^0 p\rightarrow\pi^0 p}&&\nonumber\\
 &\hspace{-3.5cm}=&\hspace{-1.9cm}\left(-\frac{\sqrt{2}}{3}v_{31}-\frac{1}{3\sqrt{2}}v_{11}\right)^2
+\left(\frac{1}{3}v_{31}-\frac{1}{3}v_{11}\right)^2,
\end{eqnarray}
\begin{eqnarray}
 \delta V_{\pi^0 p\rightarrow\pi^+ n}&&\nonumber\\
 &\hspace{-3.5cm}=&\hspace{-1.9cm}\left(-\frac{\sqrt{2}}{3}v_{31}-\frac{1}{3\sqrt{2}}v_{11}\right)
\left(\frac{1}{3}v_{31}-\frac{1}{3}v_{11}\right)\nonumber\\
&&\hspace{-1.9cm}
+\left(\frac{1}{3}v_{31}-\frac{1}{3}v_{11}\right)
\left(-\frac{1}{3\sqrt{2}}v_{31}-\frac{\sqrt{2}}{3}v_{11}\right),
\end{eqnarray}
\begin{eqnarray}
 \delta V_{\pi^+ n\rightarrow\pi^+ n}&&\nonumber\\
 &\hspace{-3.5cm}=&\hspace{-1.9cm}
\left(\frac{1}{3}v_{31}-\frac{1}{3}v_{11}\right)^2
+
\left(-\frac{1}{3\sqrt{2}}v_{31}-\frac{\sqrt{2}}{3}v_{11}\right)^2,
\end{eqnarray}
where $G_{\pi\pi N}$ is the $\pi\pi N$ loop function that incorporates the two-pion
relative momentum squared, whose analytic expression together with those of $v_{31}$ and
$v_{11}$ can be found in Ref.~\cite{Inoue:2001ip}.

\begin{figure}[htpb]
\centering
\includegraphics[scale=0.34,angle=270]{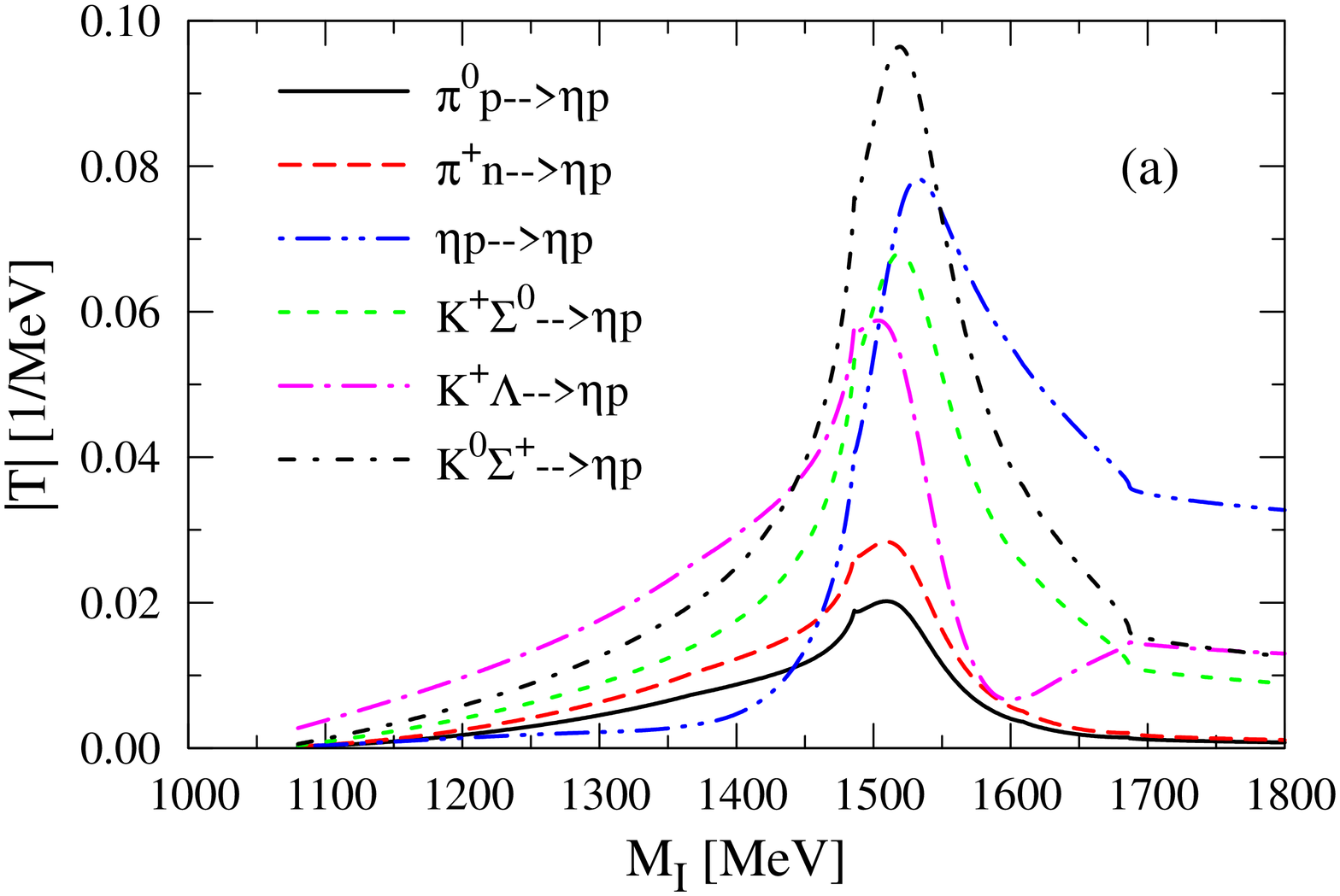}
 \includegraphics[scale=0.34,angle=270]{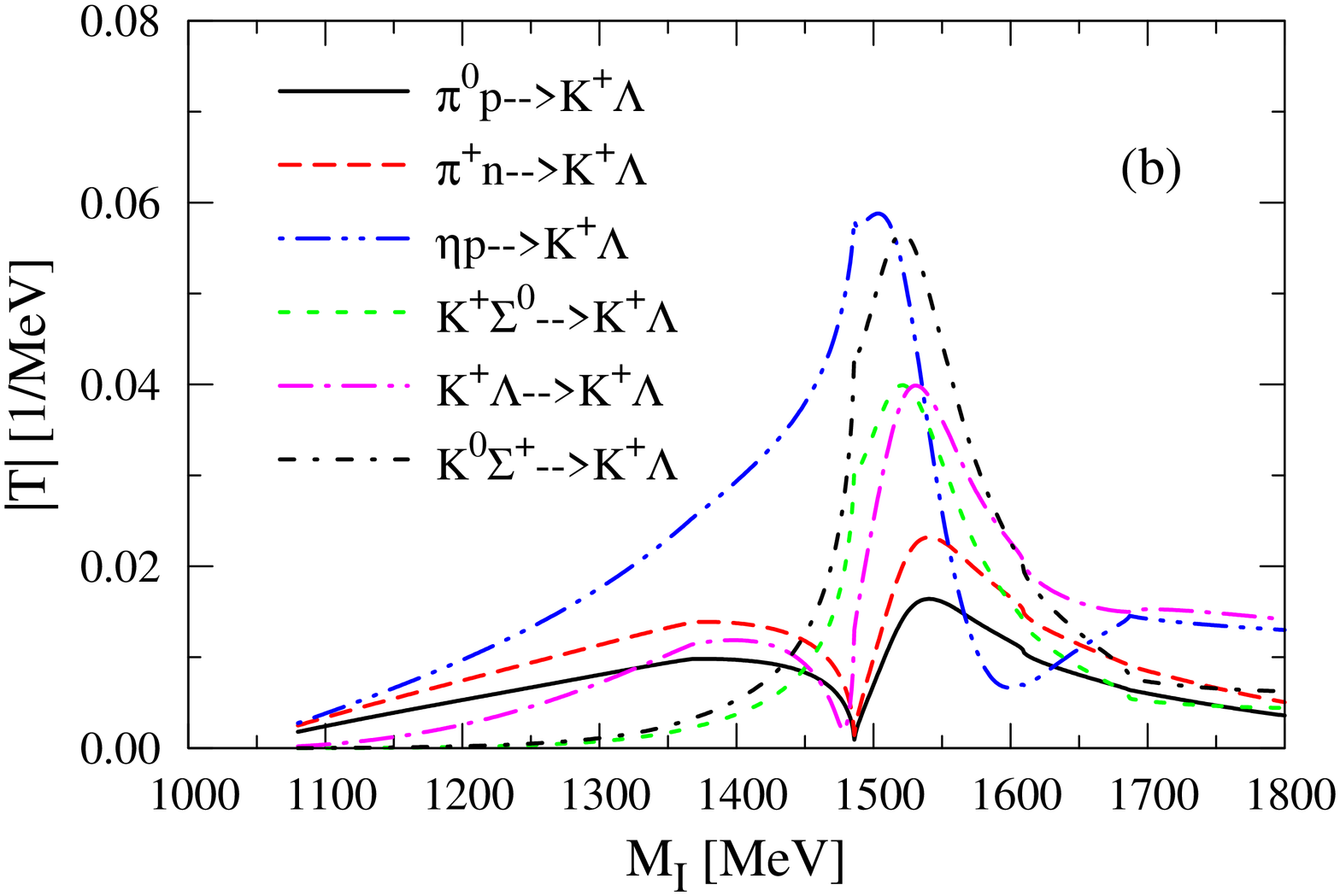}
\caption{(Color online) The moduli of the transition amplitudes in different channels
leading to the $\eta p$ and $K^+\Lambda$ final states.\label{fig:t2}}
\end{figure}

Searching for poles in the isospin 1/2 channel on the second Riemann sheet, one
finds the $N^*(1535)$ pole at $1543-i46$ MeV~\cite{Inoue:2001ip}, whose width is 
smaller than the PDG estimation of 100 $\sim$ 250 MeV but in agreement with
the BES $J/\psi\rightarrow\bar{p}\eta p$ data, $95\pm15$ MeV~\cite{Bai:2001ua}.

The moduli of the unitarized amplitudes $|T_{ij}|$ with $i$ any of the
six coupled channels and $j$ $\eta p$ or $K^+\Lambda$ are shown in Fig.~\ref{fig:t2}.
It is interesting to note that the amplitude around the $N^*(1535)$ does not behave like an usual
Breit-Wigner resonance, even at the peak position. Therefore, a pole 
simplification of this
resonance by 
\begin{equation}
T_{ij}=\frac{g_i g_j}{\sqrt{s}-M_{N^*}+i\Gamma/2}
\end{equation}
might lead to problems. We come back to this issue in Sec. IV.

\section{Reaction mechanisms of $J/\psi\rightarrow \bar{p}\eta p$ and $J/\psi\rightarrow \bar{p}K^+\Lambda$}

\begin{figure}[t]
\centering
\includegraphics[scale=0.38]{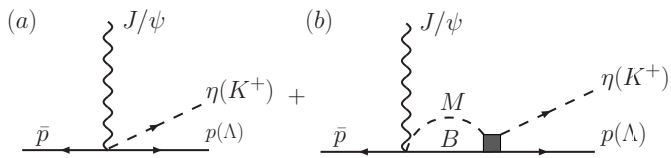}
\caption{The reaction mechanisms of $J/\psi\rightarrow \bar{p} \eta p$ and $J/\psi\rightarrow \bar{p} K^+\Lambda$ through intermediate $N^*(1535)$. For $J/\psi\rightarrow \bar{p}p\eta$ a similar diagram through
$\bar{N}^*(1535)$ has been added.\label{fig:rmechanism1}}
\end{figure}

The picture of the $N^*(1535)$ as dynamically generated from the meson-baryon
interaction has a repercussion in the mechanisms of production.  One must first produce
the relevant meson-baryon components, which upon interaction produce the resonance.
This means that
the $J/\psi$ decaying into $\bar{p}\eta p$ and $\bar{p}K^+\Lambda$ 
proceeds through the following steps: the $J/\psi$ first decays into
$\bar{p} MB$, with $MB$ being one of the six coupled channels. The rescattering of
the $MB$ pair generates dynamically the $N^*(1535)$, which then decays back into
any of the coupled channels. Such a process is illustrated in Fig.~\ref{fig:rmechanism1}.

Because the $J/\psi$ is a SU(3) singlet, its couplings 
to the $\bar{p}MB$ system can be obtained from the $D$ and $F$ terms of the lowest order chiral 
Lagrangian of Eq.~(\ref{eq:lag}). This SU(3) argument still would have $D$ and $F$ as free parameters in $J/\psi\rightarrow\bar{p}\eta p\;(\bar{p}K^+\Lambda)$. However,  the $J/\psi\rightarrow\bar{P}MB$ process 
is OZI forbidden (the $c\bar{c}$ quarks of $J/\psi$ decouple from those of the $\bar{p} MB$ system)
and it only brings into the scheme a $\vec{\sigma}\cdot\vec{\epsilon}$ operator, which is SU(3) blind.
We can then invoke SU(6) symmetry, mixing spin and flavor, to evaluate
the $\vec{\sigma}\cdot\vec{\epsilon}$ coupling with two octets of the baryons, with their
SU(3) and spin functions, and the octet of the mesons. Assuming this symmetry, the
ratio $F/D$ is fixed to the value $2/3$~\cite{PhysRevLett.13.299}, very close to the empirical value.
Because we only need the ratio $F/D$, the SU(6) symmetry provides us with the needed value.
Therefore, we take $F$ and $D$ as the empirical values up to a common constant $C$.
 The couplings are listed in Table \ref{table:jpsi},
where we have assumed $C$ to be 1 because later on we are only interested in the ratio
of the integrated cross sections, not their respective absolute values.

The $t$ matrix of the reaction mechanism of Fig.~\ref{fig:rmechanism1} can be easily written 
down (up to a global $\vec{\sigma}\vec{\epsilon}$ factor 
with $\vec{\epsilon}$ being the $J/\psi$ polarization vector) as
\begin{equation}\label{eq:t}
 t_i=\sum_{j=1}^6 D_j(\delta_{ji}+ G_j T_{j\rightarrow i})
\end{equation}
where $D_j$ is the coupling of the $J/\psi$ to channel $j$ (see
Table \ref{table:jpsi}), $G_j$ the one-baryon one-meson loop function, and
$T_{j\rightarrow i}$ the unitarized amplitude.  The corresponding invariant mass distribution for
the $J/\psi\rightarrow \bar{p}K^+\Lambda$ reaction is quite simple:
\begin{equation}
 \frac{d\Gamma}{dM_I}=\frac{M_{\bar{p}} M_\Lambda}{8\pi^3}\frac{1}{M_{J/\psi}^2} k_{\bar{p}}\tilde{k}_\Lambda|t_5|^2
\end{equation}
where 
\begin{equation}
 k_{\bar{p}}=\frac{\lambda^{1/2}(M^2_{J/\psi}, M_{\bar{p}}^2, M_I^2)}{2 M_{J/\psi}},
\end{equation}
\begin{equation}
 \tilde{k}_\Lambda=\frac{\lambda^{1/2}(M^2_I, m_K^2, M_\Lambda^2)}{2 M_I}
\end{equation}
with $M_\Lambda$ and $m_K$ being the masses of the $\Lambda$ and kaon, and $t_5$ given in Eq.~(\ref{eq:t}).
\begin{table}[t]
      \renewcommand{\arraystretch}{1.5}
     \setlength{\tabcolsep}{0.25cm}
     \centering
     \caption{The coupling of $J/\psi$ to $\bar{p}MB$ with $MB$ being one of the six coupled channels. \label{table:jpsi}}
     \begin{tabular}{ccccccc}
     \hline\hline 
       $\pi^0 p$ & $\pi^+ n$ & $\eta p$ & $K^+\Sigma^0$ & $K^+\Lambda$ & $K^0\Sigma^+$  \\\hline
       $\frac{D+F}{2f_\pi}$ & $\frac{D+F}{\sqrt{2}f_\pi}$ & $\frac{3F-D}{2\sqrt{3}f_\eta}$ & $\frac{D-F}{2f_K}$ &
       $-\frac{D+3F}{2\sqrt{3} f_K}$ & $\frac{D-F}{\sqrt{2} f_K}$\\
    \hline\hline
    \end{tabular} 
       \end{table}
       
For the reaction $J/\psi\rightarrow \bar{p}\eta p$, because it can proceed through either 
intermediate $N^*$ or intermediate $\bar{N}^*$,
one cannot derive such a simple expression. The total width for this reaction is
\begin{eqnarray}
\Gamma&=&\frac{1}{2M_{J/\psi}}\frac{1}{(2\pi)^5}\frac{M_p^2}{2}\int dE_p\int d\Omega_p\int d\omega_\eta\int d\phi_\eta\\
&&\times|t_3(N^*)+t_3(\bar{N}^*)|^2\Theta(1-A)^2)\Theta(M_{J/\psi}-E_p-\omega_\eta),\nonumber
\end{eqnarray}
where $M_p$, $E_p$, and $\Omega_p$ are the mass, energy, and solid angle of the proton,
while $\omega_\eta$ and $\phi_\eta$ are the energy of the eta and its azimuthal angle relative to
the proton, and $A$ is
\begin{equation}
A=\frac{1}{2k_p k_\eta}[(M_{J/\psi}-E_p-\omega_\eta)^2-M_p^2-k_p^2-k_\eta^2]
\end{equation}
with $k_p$ and $k_\eta$ being the moduli of the three-momenta of the proton and the eta in the
$J/\psi$ rest frame. 

The amplitude $t_3(\bar{N}^*)$ is the same as that of Eq.~(\ref{eq:t}) for
the $p\eta$ amplitude, omitting the $\delta_{ij}$ not to double count, but written as a function of the invariant mass of $\bar{p}\eta$ instead of that of $p\eta$. The consideration of $t_3(\bar{N}^*)$
accounts for the final state interaction of $\eta\bar{p}$; however, we should in
principle also care about the final state interaction of $K^+\bar{p}$, $\bar{p}p$, or
$\bar{p}\Lambda$. The $\eta\bar{p}$ interaction has been singled out because one
can have the $\bar{N}^*(1535)$ formation with $\bar{p}\eta$ as well as the $N^*(1535)$ formation 
with $p\eta$. The interactions of the other pairs are different. 
The $K^+\bar{p}$ couples strongly to
the $\bar{\Lambda}(1405)$. However, this resonance is below
the $K^+\bar{p}$ threshold, and we
are interested in the region of $K^+\Lambda$ energies around
threshold, where the invariant mass of $K^+\bar{p}$ is  far away from the $\bar{\Lambda}(1405)$ in the 
 $J/\Psi$ decay, which has 550 MeV of excess energy. 
However, one spans the region where the $\bar{\Lambda}$(1670) appears.
This resonance appears also as dynamically generated in the chiral approach
that we use for the $K^+\bar{p}$ ($K^- p$) interaction~\cite{Oset:2001cn}, and, because of that, we 
take this interaction into account.
We can use similar arguments for the  $\bar{p}p$ and $\bar{p}\Lambda$ interactions, which will have
relatively large invariant masses. In these cases the potential energy is small compared
to the kinetic energy and accordingly the wave function diverts little from the plane wave,
thus, barely modifying the production amplitudes~\cite{Gold64,Watson:1952ji,Hanhart:2003pg}. 

\section{Results and discussions}
\subsection{Comparison with the data}
\begin{figure}[t]
\centering
\includegraphics[scale=0.34,angle=270]{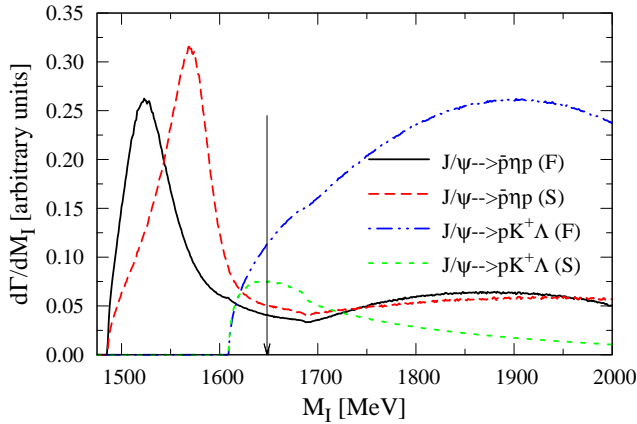}
\caption{(Color online) Invariant mass distributions of $J/
\psi\rightarrow \bar{p}\eta p$ and $J/\psi\rightarrow \bar{p}K^+\Lambda$.
The theoretical results labeled ``F'' are obtained using the full U$\chi$PT amplitudes
containing the vertex and $\pi\pi N$ loop corrections, while the ones labeled ``S''
are obtained by using the  amplitudes without these two corrections and with the
readjusted subtraction constants by fitting the $\pi^-p\rightarrow K Y$ data~\cite{Doring:2008sv}.
\label{fig:inv1}}.
\end{figure}

The invariant mass distributions for the reactions
$J/\psi\rightarrow \bar{p}\eta p$ and $J/\psi\rightarrow \bar{p}K^+\Lambda$ 
are shown in Fig.~\ref{fig:inv1}. As argued in the previous section, we have used the same
$D$ and $F$ coefficients as the couplings of the pseudoscalars to the baryon octet:
$D=0.795$ and $F=0.465$~\cite{Gold64, Close:1993mv,Borasoy:2007ku}. In the figure, 
the curves labeled with an ``F'' are obtained with the full U$\chi$PT amplitudes as described above,
while the curves labeled with an ``S'' are obtained with the amplitudes without
the $\pi N\rightarrow \pi\pi N$ and vertex corrections as explained below.
It is seen that the differences between the results obtained with the full amplitudes and those
with the amplitudes without the $\pi\pi N$ channel and vertex corrections are rather large beyond $\sim$1650 
MeV. This should not worry us much because we are only interested up to this energy. Below
this energy, the invariant mass distributions peak at slightly different energies, but this
does not change the integrated cross section a lot.

Now we are in a position to compare the theoretical ratio of the two integrated cross sections with
the data. However, our model for the dynamical generation of the
$N^*(1535)$ is reliable only up to $\sim$1650 MeV. Thus, we can only compare
the integrated decay width up to this energy. The experimental ratio is estimated to be
\begin{eqnarray}\label{eq:data1}
 R_\mathrm{exp.}&=&\frac{\Gamma(J/\psi\rightarrow \bar{p}K^+\Lambda)}
{\Gamma(J/\psi\rightarrow  \bar{p}\eta p)}\\
&=&
\frac{(0.89\pm0.16\times10^{-3})\times 10\%}{(2.09\pm0.18\times10^{-3})\times 31\%}
\approx0.14\pm0.04.\nonumber
\end{eqnarray}
The numbers $0.89\pm0.16\times 10^{-3}$ and $2.09\pm0.18\times10^{-3}$
are the branching ratios for the $J/\psi$ decaying into $\bar{p}K^+\Lambda$ and
$\bar{p}\eta p$~\cite{Yao:2006px}. 
The fraction $10\%$ of the
strength of $J/\psi\rightarrow \bar{p}K^+\Lambda$ up to $M_I=1650$ MeV 
is estimated by studying
the $J/\psi\rightarrow\bar{p}K^+\Lambda$ experimental spectrum (see Fig.~9b of Ref.~\cite{Yang:2005ej}). 
The number $31\%$ is the estimated fraction of the 
amount of $J/\psi\rightarrow\bar{p}\eta p$ up to the same energy (see Fig.~8 of Ref.~\cite{Bai:2001ua}).

It should be noted that the above ratio has been obtained by using only the raw data to 
avoid uncertainties related to the further treatments of the data. On the other hand, 
 using the results of the partial wave analyses of  Refs.~\cite{Bai:2001ua} and \cite{Yang:2005ej}, 
 the ratio is estimated to be 
\begin{eqnarray}\label{eq:data2}
 R_\mathrm{exp.}&=&\frac{\Gamma(J/\psi\rightarrow \bar{p}N^*\rightarrow \bar{p}K^+\Lambda)}
{\Gamma(J/\psi\rightarrow \bar{p}N^*+p\bar{N}^*\rightarrow \bar{p}\eta p)}\\
&=&\frac{(0.89\pm0.16\times10^{-3})\times (15\sim22)\%}{(2.09\pm0.18\times10^{-3})\times(56\pm15)\%}
\approx0.14^{+0.15}_{-0.07},\nonumber
\end{eqnarray}
which is the ratio fitted to obtain the $N^*(1535)$ coupling
constant to $K\Lambda$ 
in Ref.~\cite{Liu:2005pm}. We note that the ratios obtained either way are consistent with
each other, albeit with large uncertainties.

On the other hand, our theoretical ratio of the integrated cross sections 
from the respective thresholds up to
$M_I=1650$ MeV (using the full amplitudes) is
\begin{equation}
R_\mathrm{th}=\frac{\Gamma(J/\psi\rightarrow \bar{p}N^*\rightarrow \bar{p}K^+\Lambda)}
{\Gamma(J/\psi\rightarrow \bar{p}N^*+p\bar{N}^*\rightarrow \bar{p}\eta p)}
=0.16^{+0.06}_{-0.04},
\end{equation}
which is in reasonable agreement with the experimental ratio determined either way. 
The theoretical uncertainties are estimated by slightly changing the $F/D$ ratio appearing
in the $J/\psi$ couplings to $\bar{p}MB$ by $5\%$. 

It is interesting to note that the theoretical ratio is obtained by assuming SU(6) symmetry 
for the $J/\psi$ to $\bar{p}MB$ couplings and by assuming, for the reasons given above, that the $D$ and $F$ coefficients are the same
as those appearing in the Yukawa couplings of one pseudoscalar to the octet of  baryons, up to a global
constant. The agreement with the data supports these assumptions.

Another source of inherent theoretical uncertainties comes from the consideration of
the $\pi\pi N$ channel,  the vertex correction, and the freedom one has in the values of
the subtraction constants. Following Ref.~\cite{Doring:2008sv}, 
we assess these uncertainties by removing the contribution of
the $\pi\pi N$ channel and the vertex corrections and adjusting the subtraction constants
to fit the $\pi^-p\rightarrow K Y$ cross-section data at higher energies. The corresponding
invariant mass distributions obtained this way are shown in Fig.~\ref{fig:inv1}, the curves labeled
with an ``S''.
It is seen that the $\eta p$ peak position is moved to slightly higher energies by
$\sim$30 MeV, while the ratio of the two integrated cross sections is reduced to
$\sim$0.11. Combining the results with the full and the ``simplified'' U$\chi$PT amplitudes, 
we arrive at the theoretical ratio
\begin{equation}
R_\mathrm{th}=0.135\pm0.06,
\end{equation}
where the central value is an average of the ratios obtained with the full amplitudes and the simplified amplitudes, and the dispersion incorporates both the uncertainties in the U$\chi$PT amplitudes and
those in the $D$ and $F$ coefficients.

\begin{figure}[t]
\centering
\includegraphics[scale=0.34,angle=270]{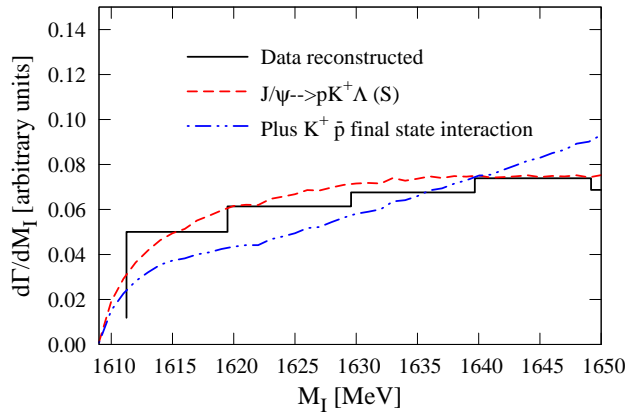}
\caption{(Color online) The invariant mass distributions of the
 $J/\psi\rightarrow \bar{p}K^+\Lambda$ reaction with and without the $K^+\bar{p}$ final 
state interaction~\cite{Oset:1997it}.
\label{fig:kpluspbar}}
\end{figure}

Above we have shown that the two experimental ratios Eq.~(\ref{eq:data1}) and Eq.~(\ref{eq:data2})
are consistent with each other. However, there is a caveat in our comparison with these two numbers.
Our above comparison with the experimental ratio of Eq.~(\ref{eq:data2}) is fine, because this ratio
is obtained solely from $N^*$ contributions as explained by the corresponding experimental analysis. Our
comparison with the experimental ratio of Eq.~(\ref{eq:data1}), on the other hand, is not very consistent because in this
case we have to include the $K^+ \bar{p}$ final state interaction in the same way we included
the $\eta\bar{p}$ interaction. The $K^+\bar{p}$ interaction has been studied extensively in
unitary chiral theories and is well understood around the $K^+\bar{p}$ threshold and, to a lesser degree,
around the $\bar{\Lambda}(1670)$ peak position. For energies beyond the $\bar{\Lambda}(1670)$ peak
position, it is less constrained and the comparison with data is only qualitative~\cite{Oset:2001cn}. 
Despite all the uncertainties of the $K^+\bar{p}$ interaction, it is still interesting to see how
adding this part will change the scenario. Following  a procedure similar to that of including the
$\eta\bar{p}$ contribution and using the same argument to obtain the couplings of the $J/\psi$ to
the ten channels coupling to $K^+\bar{p}$, we find that adding the $K^+\bar{p}$ final state interaction
only changes our calculated ratio by $\sim$$10\%$ percent, which is smaller than the theoretical errors.
The corresponding invariant mass distributions are shown in Fig.~\ref{fig:kpluspbar}, in comparison
with the results obtained without including the $K^+\bar{p}$ final state interaction---the curve denoted by (S) in Fig.~3. The experimental numbers are obtained by multiplying by phase space  
the numbers shown in Fig.~9 of Ref.~\cite{Yang:2005ej}.

We can conclude from the above discussion that the theoretical ratio is rather stable despite
the relatively large uncertainties of the $K^+\bar{p}$ final state interaction. This is true partly
because we confine ourselves to $K^+\Lambda$ center of mass energies below $\sim$1650 MeV where the $K^+\bar{p}$ interaction is relatively
well constrained due to the three-body phase space.

\subsection{Couplings constants in different models}

It is instructive to compare the ratio of the $N^*(1535)$ effective couplings 
obtained in the unitary chiral approach of Ref.~\cite{Inoue:2001ip} and that obtained in Ref.~\cite{Liu:2005pm}, because both models describe the $J/\psi$ decay data. 
Using the numbers from Ref.~\cite{Inoue:2001ip}, one has
$|g_{N^*(1535) K\Lambda}|=0.92$ and $|g_{N^*(1535)\eta N}|=1.84$ obtained from the residues of the $T$ amplitudes at the pole position on the complex plane.
On the other hand, from the same study, one finds
$|g_{N^*(1535) K\Lambda}|=1.28$ and $|g_{N^*(1535)\eta N}|=1.77$ through a Breit-Wigner fit of the real energy
scattering amplitudes. Thus,
we obtain
\begin{equation}
R=\frac{|g_{N^*(1535) K\Lambda}|}{|g_{N^*(1535)\eta N}|}=0.5\sim0.7.
\end{equation}
This is a factor of two smaller than the one obtained in Ref.~\cite{Liu:2005pm},
$1.3\pm 0.3$, from the comparative study of the reactions studied in the present work,
 and
is slightly smaller than the range of $0.8\sim2.6$ given in Ref.~\cite{Penner:2002md}.

The relatively large discrepancy between the phenomenologically determined $R$ and
the U$\chi$PT ones reveals a fundamental difference in these two different descriptions of
resonances, particularly in the region far from the resonance peak position, which is relevant to the
present study. In the phenomenological description  one often adopts a Breit-Wigner-like formula to describe
the distribution of a resonance,
\begin{equation}\label{eq:spole}
\frac{\tilde{g}_i \tilde{g}_j}{(S-M^2)+i M\Gamma(s)},
\end{equation}
where $\tilde{g}_i$, $\tilde{g}_j$ are the coupling constants of the resonance to channels $i$ and $j$, $M$ is the mass of
the resonance, and $\Gamma(s)$ is the width of the resonance, which incorporates the explicit energy dependence. 
This is the type of amplitudes used in Ref.~\cite{Liu:2005pm} to describe the $J/\psi$ decay processes.  

This approximation 
assumes that the resonance's shape in different channels is the same, while the only difference comes from the coupling constants. This could be a valid approximation in many cases, but it is not true
in the present case, as can be clearly seen from Fig.~\ref{fig:t2}. There, one can easily
see that even around the resonance peak position the shapes in different channels are not proportional to
each other. The deviations become even larger at the $K^+\Lambda$ threshold. 
The dynamics of coupled channels is mostly responsible for that \cite{Kaiser:1995cy,Kaiser:1996js,Nieves:2001wt,Inoue:2001ip},
and particularly the large coupling of the resonance to the $\eta N$ channel close to
threshold. This  particular behavior of the resonance might explain
the relatively large discrepancy between the coupling constants obtained in different methods, though
they both describe the data. 

One may well conclude that the coupling constants determined from chiral unitary theory and those determined
from phenomenological studies cannot be directly compared: They only have meanings inside the framework where
they are deduced, at least quantitatively.  This has also been pointed out recently in Ref.~\cite{Doring:2008sv}.

Of course, one of the aims of the present work is to show consistency of the present
two $J/\psi$ decay reactions with the idea of the $N^*(1535)$ resonance as being dynamically generated in
the chiral unitary approach, and we see that indeed the U$\chi$PT picture is consistent with
the BES data. Note that we have not fitted the experimental
data to obtain the $N^*\Lambda K$ couplings, as is the case in Ref.~\cite{Liu:2005pm}. We have
used the results of Ref.~\cite{Inoue:2001ip} in the context of the two $J/\psi$ reactions and
have found consistency with the data, with some uncertainties tied to the nature of these two
reactions beyond the dynamics of the meson-baryon interaction, as we have explained above. We found consistency with the data even if the
U$\chi$PT picture produces
a coupling for $N^*K\Lambda$ different than that in Ref.~\cite{Liu:2005pm}, but we also have mentioned that the meaning of the couplings is not exactly the same because of the different shapes used for the energy
dependence of the amplitudes. Furthermore, the approach followed here does not make any explicit use of
the $g_{N^*(1535)K\Lambda}$ coupling because we used the full $MB$ amplitudes. These amplitudes are different from
the simple pole approximation that one would obtain extrapolating the form of Eq.~(\ref{eq:spole}) with constant
$\Gamma$ to higher energies. 

A further remark concerning the meaning of the couplings: The one
obtained in Ref.~\cite{Inoue:2001ip} comes from the residue of the $N^*$ pole, while the one from Ref.~\cite{Liu:2005pm}
comes from fits to data around the $K\Lambda$ threshold. The former
is a measure of the strength of the $K\Lambda$ component in the $N^*(1535)$ wave function and plays a role
in the determination of the properties of the $N^*(1535)$. For instance, if one wishes to determine the helicity amplitudes of the $N^*(1535)$ as done in Ref.~\cite{Jido:2007sm}, it is the residue of the $N^*$ pole that must be used in the calculation.

\section{Summary}
We have studied the $J/\psi\rightarrow \bar{p}\eta p$ and $
J/\psi\rightarrow \bar{p}K^+\Lambda$ reactions, more specifically, the ratio of the
integrated cross sections, using the unitary chiral approach. 
The unitary chiral approach, which generates the $N^*(1535)$ dynamically, can
describe the data reasonably well. This was despite the fact that the coupling of the
$N^*(1535)$
to the $K\Lambda$ channel is different from the one obtained in the empirical study of the present reactions
in Ref.~\cite{Liu:2005pm}, but it was clarified that the concepts are different. The couplings of the chiral unitary approach come from the residues of the amplitudes at the $N^*$ pole, while the $N^*\Lambda K$
coupling obtained in Ref.~\cite{Liu:2005pm} comes from a fit to the data close to the  $K\Lambda$
threshold assuming a certain shape for the $N^*$ dominated amplitude.  
Furthermore,
it is interesting to note that although the couplings obtained in different ways are quantitatively different,
they both indicate that the $N^*(1535)$ wave function contains a large $s\bar{s}$ component.

Certainly, the $N^*(1535)$ may in fact be a mixture of three-quark component and five-quark (meson-baryon) component,
as suggested by the studies of Ref.~\cite{Hyodo:2008xr}. This may slightly change  the numbers
obtained in this work, but the main conclusions, taking into account both theoretical and experimental
uncertainties, will remain the same.

\section{Acknowledgments}
We would like to thank Pedro Gonzalez for useful discussions.
This work is partly
supported by FIS2006-03438, by the National Natural Science Foundation
of China and the Chinese Academy of Sciences under
Project KJCX3-SYW-N2, by the
Generalitat Valenciana,  and by the EU Integrated
Infrastructure Initiative Hadron Physics Project under Contract RII3-CT-2004-506078.

\bibliographystyle{apsrev}
\bibliography{jpsi_nstar}
\end{document}